**Ferrimagnetic Skyrmions in a Tetragonal $Mn_{1.9}Co_{0.1}Sb$ Single Crystal at Room Temperature**

Huanhuan Zhang, YaJiao Ke*, Weiwei Wang, Lingyao Kong, Lin Chen, Sheng Qiu, Jialiang Jiang, Yongsen Zhang, Youhong Peng, Yaodong Wu*, Mingliang Tian, Haifeng Du*, and Jin Tang*

Huanhuan Zhang, Weiwei Wang, Lingyao Kong, Jialiang Jiang, Yongsen Zhang, Mingliang Tian, Jin Tang

State Key Laboratory of Opto-Electronic Information Acquisition and Protection Technology, School of Physics, Anhui University, Hefei, 230601, China

Email: jintang@ahu.edu.cn

Yajiao Ke, Lin Chen

School of Physics and Mechanics, Wuhan University of Technology, Wuhan 430000, China

Email: keyajiao@whut.edu.cn

Sheng Qiu, Youhong Peng, Yaodong Wu, Mingliang Tian, Haifeng Du, Jin Tang

Anhui Provincial Key Laboratory of Low-Energy Quantum Materials and Devices, High Magnetic Field Laboratory, HFIPS, Chinese Academy of Sciences, Hefei 230031, China

Email: duhf@hmfl.ac.cn

Yaodong Wu

School of Physics and Materials Engineering, Hefei Normal University, Hefei, 230601, China

Email: wuyaodong@hfnu.edu.cn

**Abstract**

The development of room temperature small-sized ferrimagnetic skyrmion materials is significant for topological spintronic device applications. As a room temperature ferrimagnetic material, the tetragonal $Mn_{1.9}Co_{0.1}Sb$ crystal exhibits multiple phase transitions, including spin reorientation transitions. However, the magnetic spin textures and their evolution mechanisms during magnetic phase transitions in $Mn_{1.9}Co_{0.1}Sb$ crystals remain unexplored. Using Lorentz transmission electron microscopy, we discovered and verified dipolar skyrmion behavior and its magnetic evolution at room temperature. We established a stable phase diagram of magnetic textures as functions of temperature and magnetic field, while also investigating the evolution mechanisms of spin textures across multiple temperature-induced magnetic phase transitions. Through micromagnetic simulations, a ferrimagnetic configuration with in-plane ferromagnetic coupling and interlayer antiferromagnetic arrangement was established, which stands in contrast to synthetic ferrimagnetic/antiferromagnetic systems that exhibit interlayer antiferromagnetic coupling via the Ruderman-Kittel-Kasuya-Yosida (RKKY) interaction. We determined that the intrinsic frequency of ferrimagnetic skyrmions can reach the THz regime due to strong interlayer antiparallel exchange interactions. These findings highlight the diversity of room temperature ferrimagnetic skyrmion regulation behaviors in $Mn_{1.9}Co_{0.1}Sb$ and their dynamic evolution characteristics, opening new avenues for developing novel spintronic devices with enhanced functionalities capable of operating under ambient conditions.

Keywords:

magnetic skyrmions, ferrimagnetic, $Mn_{1.9}Co_{0.1}Sb$, nanostructured cells, spintronics

## 1. Introduction

Magnetic skyrmions are nanoscale magnetic structures characterized by non-trivial topological order, representing a promising platform for developing next-generation memory and logic devices. [1-8] Early investigations primarily concentrated on ferromagnetic systems, where skyrmions were first identified in the metallic alloy MnSi via neutron scattering, [9] and subsequently directly imaged in $Fe_{0.5}Co_{0.5}Si$ using Lorentz transmission electron microscopy (LTEM).[10] Further advances in materials like FeGe[11-13] and CoZnMn[14, 15] enabled skyrmion manipulation near room temperature conditions. The emerging topology-related electromagnetic properties of magnetic skyrmions enable their potential for racetrack memory applications and neuromorphic computing. [7, 16-18] However, the practical implementation of ferromagnetic skyrmions faces significant challenges due to their skyrmion Hall effects, [19-21] and slow dynamic frequency. [22, 23]

In this context, ferrimagnetic skyrmions have emerged as a prominent candidate due to their unique coupling mechanism of antiparallel and non-equivalent magnetic moments. [24-26] By partially canceling out the net magnetization, ferrimagnetic systems can reduce skyrmion Hall effects. [27, 28] The skyrmion Hall effect in synthetic ferrimagnets should not be judged solely on the absolute value of the net magnetic moment, but rather on the effective topological number $\bar{Q}$. [29] Skyrmion Hall effects are completely canceled when $\bar{Q} = 0$, which corresponds to the synthetic antiferromagnets. The skyrmion Hall angle is approximately proportional to $\bar{Q}$ when it is in the limit approaching 0. Thus, ferrimagnets with $0<|\bar{Q}|<1$ could possess reduced skyrmion Hall effects compared with ferromagnets with $|\bar{Q}|=1$. Moreover, the antiparallel exchange interaction could contribute to a high-frequency Terahertz dynamic response for high-speed information processing. [30-32] Nevertheless, existing ferrimagnetic materials capable of sustaining stable and tunable nanoscale skyrmions at room temperature have been reported few material systems. [33-35]

$Mn_{2-x}Co_xSb$ crystal with a tetragonal lattice is reported as a room temperature ferrimagnetic material with multiple magnetic phase transformations, including paramagnetic to ferrimagnetic transformation and the spin reorientation transition. [36] Previous studies about $Mn_{1.9}Co_{0.1}Sb$ including magnetoresistance and magnetocaloric effects. [37-39] A previous study shows stripe domains with widths of approximately 195 nm in $Mn_{1.9}Co_{0.1}Sb$, [40] but the direct observation of magnetic skyrmions and their magnetic evolution under magnetic phase transformations has not been explored yet.

Based on those foundations, our current study demonstrates the first experimental verification of room temperature dipolar skyrmions in $Mn_{1.9}Co_{0.1}Sb$ ferrimagnets through systematic LTEM investigations. Due to the central symmetry of the tetragonal ferrimagnet, coexisting skyrmions with opposite winding directions were also observed at room temperature, a phenomenon further verified through micromagnetic simulations. These simulations reveal magnetic structure transformations during temperature-induced phase transitions, at the same time provide compelling evidence supporting the high resonance frequencies and ultrafast response advantages of ferrimagnets. This research not only elucidates the formation mechanism and temperature-driven evolution of topological magnetic structures in $Mn_{1.9}Co_{0.1}Sb$ ferrimagnets but also establishes a novel materials platform for developing next-generation spintronic devices.

## 2. Results and Discussions

### 2.1. Analysis of the $Mn_{1.9}Co_{0.1}Sb$ Crystal Structure and Phase Composition

The compound $Mn_2Sb$ with the tetragonal $Cu_2Sb$ structure is ordered ferrimagnetically below 550 K and exhibiting spin reorientation transition at 240 K. Mn atoms occupy two distinct crystallographic positions, the spins of the two kinds of Mn atoms, namely Mn(I) and Mn(II), align antiferromagnetically with different magnetic moments of 2.1 μB and 3.9 μB, [41, 42] respectively. Doping $Mn_2Sb$ with transition metals such as Co, V, Cr, Cu, and Zn can also be used to control the magnetic properties. [40–45] Specifically, these dopants occupy specific lattice sites, inducing lattice contraction. This, in turn, alters the exchange interaction energy, magnetic dipole-dipole interaction energy, and magnetocrystalline anisotropy energy, thereby bringing about a rich array of magnetic phase transitions. $Mn_{1.9}Co_{0.1}Sb$ still maintains the ferrimagnetic structure. However, the ferrimagnetic transition temperature is reduced, while the spin reorientation transition temperature exhibits a slight increase. [36] Herein, we prepared $Mn_{1.9}Co_{0.1}Sb$ single crystals using the melt-cooling method. **Figure 1**(a) displays the crystal model of $Mn_{1.9}Co_{0.1}Sb$, which features a tetragonal lattice (space group P4*/nmm*). [43-48] To analyze the elemental composition, we utilized an Energy-Dispersive Spectrometer (EDS), as shown in Figure S1. We conducted X-Ray Diffraction (XRD) experiments on the plane perpendicular to the c-axis to analyze the crystal structure of the synthesized $Mn_{1.9}Co_{0.1}Sb$ crystal. Figure 1(b) shows well-defined reflection peaks that can be unambiguously indexed to the (00*l*) plane of the tetragonal phase, confirming both the high quality and proper orientation of the grown $Mn_{1.9}Co_{0.1}Sb$ single crystal. Inset is a photograph of the grown $Mn_{1.9}Co_{0.1}Sb$ single crystal showing the (00*l*) plane. Furthermore, powder XRD patterns obtained from ground samples (Figure S2) indicate the absence of any impurity phases, with calculated lattice parameters are $a = b = 4.08$ Å and $c = 6.39$ Å.

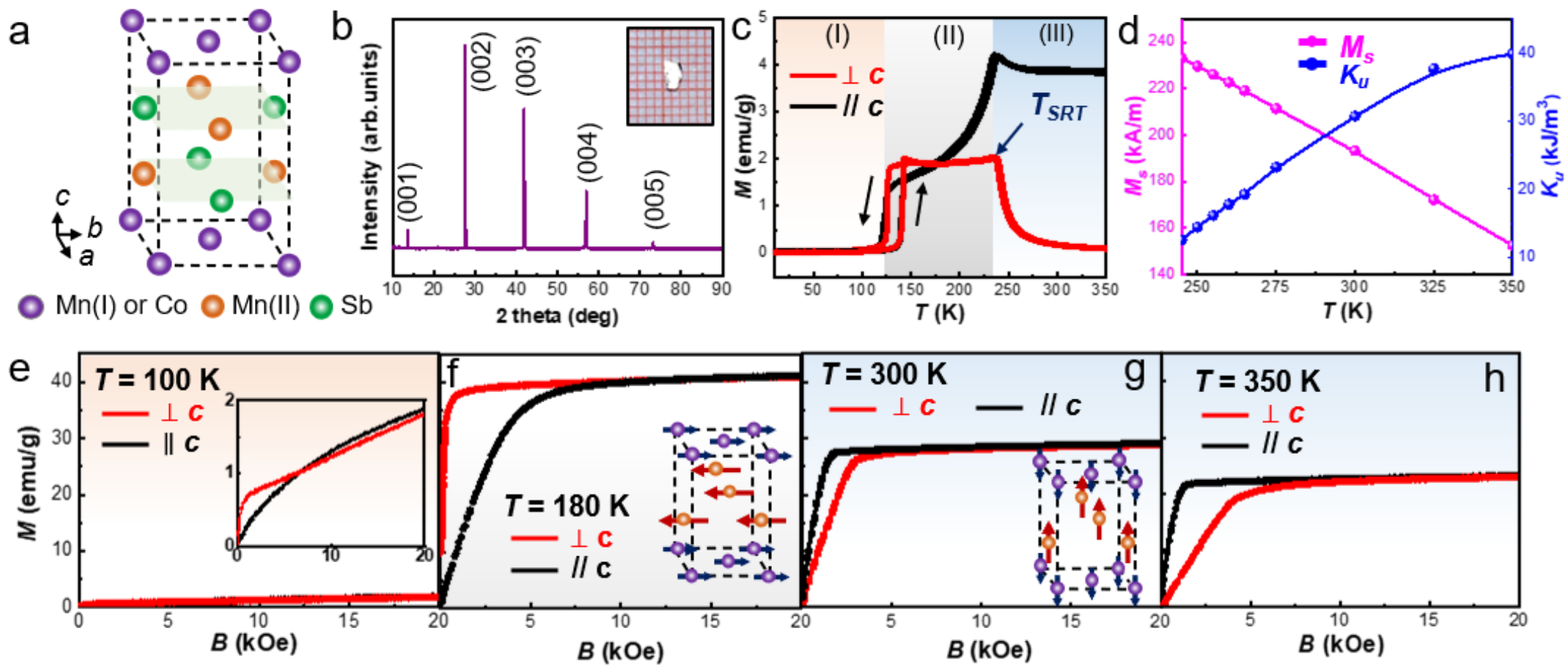

**Figure 1**. Structure and magnetic properties of $Mn_{1.9}Co_{0.1}Sb$ single crystal. a) Schematic diagram of the $Mn_{2-x}Co_xSb$ crystal structure. b) The XRD pattern of the tetragonal $Mn_{1.9}Co_{0.1}Sb$ single crystal recorded at room temperature. Regions I-III are temperature ranges defined by their distinct magnetic characteristics. c) $M$ ($T$) curves measured under a field of 100 Oe for $Mn_{1.9}Co_{0.1}Sb$ alloy, including $M_{\perp}(T)$ (red curve) and $M_{//}(T)$ (black curve). d) Magnetization $M_s$ and magnetic anisotropy $K_u$ as a function of $T$. e)-h) Isothermal magnetization curves of $Mn_{1.9}Co_{0.1}Sb$ at 100 K (e), 180 K (f), 300 K (g), and 350 K (h), left vertical axis in emu/g, right vertical axis in μB/(Mn/Co).

To analyze the macroscopic magnetic properties of the $Mn_{1.9}Co_{0.1}Sb$ crystal, we used a Physical Property Measurement System (PPMS) to observe the *M-H* and *M-T* curves for analysis. Figure 1(c) shows $M_{\perp}(T)$ (field perpendicular to c-axis) and $M_{//}(T)$ (field parallel to c-axis) under an applied magnetic field of 100 Oe, with the black arrows indicating the cooling and heating processes, respectively. A significant discontinuity in magnetization is observed at ~125 K, indicating a transition potentially originating from an antiferromagnetic to ferrimagnetic phase transition or reorientation due to strong magnetic anisotropy. A second transition occurs at ~245 K, where the decrease in $M_{\perp}(T)$ and the increase in $M_{//}(T)$ during heating suggest the induction of the spin reorientation (SRT), in which the moments of Mn (I) and Mn (II) atoms rotate from the in-plane to out-of-plane direction. Based on the magnetic properties described above, we divided the temperature range into three regions: Region I ($T <$ 125 K), Region II (125 K $< T <$ 245 K), and Region III ($T >$ 245 K), to facilitate subsequent in-depth analysis of the magnetic properties in each region. To investigate the SRT and its interactions observed in the $M$ ($T$) curves, we first calculated the uniaxial magnetic anisotropy ($K_u$) and saturation magnetization ($M_s$) as functions of temperature in the temperature range of 245 to 350 K, with the magnetic field applied along the in-plane direction. Figure 1(d) shows that $K_u$ increases with increasing temperature, while $M_s$ decreases, which is consistent with the trend of magnetization rotating direction.

We systematically analyzed the evolution characteristics of isothermal magnetization curves ($M_\perp(H)$) and ($M_{//}(T)$) at different temperatures in Figure 1(e)-(h) and Figure S3, when the external field is applied along the easy magnetization axis, higher energy is required to overcome the crystalline anisotropy for magnetic moment reversal, manifested as an increased area enclosed by the hysteresis loops. At 100 K (Region I, Figure 1(e) and Figure S3(a)-(b)), both $M_\perp(H)$ and $M_{//}(T)$ exhibit nearly linear increases without distinct saturation plateaus. The inset shows an enlarged view of the low-field region. When the temperature rises to 180 K (Figure 1(f) and Figure S3(c)-(d)) within the ferrimagnetic phase (Region II, 125-240 K), $M_\perp(H)$ develops pronounced saturation plateaus, and the nearly constant saturation field confirms that the easy magnetization axis is perpendicular to the $c$-axis in this range. Upon spin reorientation (Region II to Region III), the easy axis changes from perpendicular (Figure 1(f)) to parallel (Figure 1(g)) with respect to the $c$-axis. As shown in Figure 1(g), 1(h) and Figure S3, the magnetic anisotropy energy gradually increases with temperature, as further illustrated in Figure 1(d). By sweeping magnetic field between +5 kOe and -5 kOe, we obtained a nearly zero remanence in the hysteresis loop (as shown in Figure S4). This is attributed to the dominant role of the demagnetization energy at low fields, which promotes the formation of stripe domains in the sample (Figure 2), thereby resulting in a negligible net magnetization. The insets in Figure 1(f) and 1(g) schematically show the magnetic structures of Mn (I) and Mn (II) atoms in their respective temperature regions (background colors match Figure 1(c)). In contrast, the saturation field along the *c*-axis remains stable throughout the entire temperature range (Region III), demonstrating that the *c*-axis persistently serves as the easy magnetization axis.

**2.2. Observation of Dipolar Skyrmions in $Mn_{1.9}Co_{0.1}Sb$ Lamella**

We prepared a [001]-oriented $Mn_{1.9}Co_{0.1}Sb$ layered structure sheet with a thickness of approximately 140 nanometers using a focused ion beam (FIB) dual-beam system. The magnetic field-driven evolutions of magnetic structures in the nanostructured cells are investigated via Lorentz TEM. [49] **Figure 2**(a) illustrates the evolution of the magnetic structure in the prepared layered sample under applied fields ranging from 0 mT to 170 mT at 260 K. At zero magnetic field, typical stripe domains with an average width of ~250 nm initially appear, stabilized by the interplay between perpendicular magnetocrystalline anisotropy and dipole-dipole interactions. Upon application of a perpendicular out-of-plane magnetic field, the stripe domains gradually contract and coexist with magnetic structures featuring closed vortex-like domain walls (observed at 92 mT). These localized magnetic configurations—stabilized by the interplay among uniaxial anisotropy, the Zeeman energy, and demagnetization effects—are collectively referred to as "dipolar skyrmions".[50-52] As the magnetic field strength increases further, the system undergoes a complete transition into a dipolar skyrmion state (evident at 150 mT and 170 mT). Subsequently, when the magnetic field is further increased to saturation, the size of the magnetic skyrmions decreases, eventually vanishing in the ferromagnetic state. Overall, the balance between perpendicular magnetocrystalline anisotropy and dipole-dipole interactions under zero magnetic field promotes the formation of stripe domains. Upon applying an external perpendicular magnetic field, the Zeeman energy drives the collapse of the stripe domain walls, leading to the formation of skyrmions with integer topological charge (see Figure S5). These dipolar skyrmions vanish as the magnetic field is further increased.

Figure 2(b) displays the temperature-dependent evolution of stripe magnetic domain configurations under zero magnetic field. Stripe domains represent the spontaneous magnetic ground state in the absence of a magnetic field (Figure S6). In the low-temperature 150 K, where a significant magnetization reduction occurs, FIB prepared samples exhibit well-defined in-plane stripe domains (indicated by purple arrows) through LTEM characterization. As the temperature rises to near the $T_{SRT}$ (at 245 K), the easy axis gradually rotates from inplane to outplane, resulting in the sparse stripes state where inplane domains (indicated by brown arrows) coexist with stripe domains. At room temperature (at 293 K), the easy axis is completely oriented out-of-plane, forming distinct stripe domains. As the temperature increases (at 330 K), the stability of the stripe domains decreases due to thermal excitation, leading to wider spacing between magnetic domains, as shown in Figure 2(b).

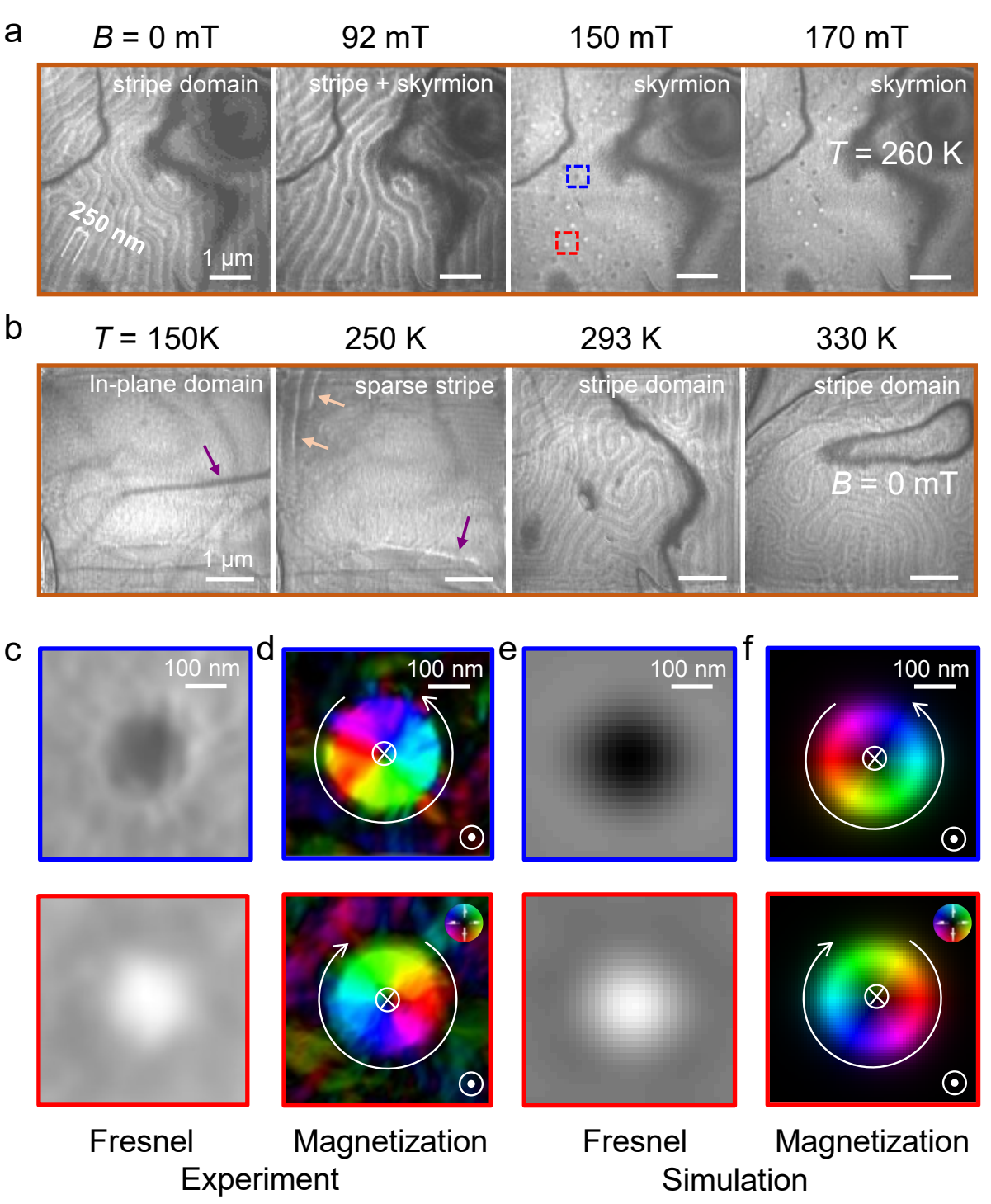

**Figure 2**. Observation of dipolar skyrmions near room temperature. a) Magnetic evolution from zero-field stripe domains driven by out-of-plane magnetic fields at $T = 260$ K. Scale bar: 1 μm. b) Evolution of zero-field stripe domains at different temperatures. c) Two skyrmions with opposite helicities observed in Lorentz experiments. The defocused distance is -2 mm. d) Representative in-plane magnetization mapping of two skyrmions retrieved from TIE analysis. e) Simulated Fresnel contrast of skyrmions based on measured magnetic parameters. f) In-plane spin configuration of simulated skyrmions.

In tetragonal ferrimagnets, the centrosymmetric nature precludes the existence of chiral Dzyaloshinskii-Moriya interactions, allowing the stable coexistence of two skyrmions with opposite winding directions (Figure 2(c)-2(f)). As shown in Figure 2(c), counterclockwise skyrmions exhibit defocused dark-field contrast (blue rectangles) due to electron wave phase modulation, while clockwise skyrmions display bright-field contrast (red rectangles), directly reflecting the chirality-dependent magnetization orientation. The Transport Intensity Equation (TIE) phase reconstruction (Figure 2(d)) further resolves the topological charge signs and boundary structures of skyrmions by analyzing the rotational direction of localized phase vortices. [53] By applying experimentally measured magnetic parameters to simulations, the simulated Fresnel images (Figure 2(e)) reproduce the inversion of bright-dark contrast, and the simulated in-plane magnetization distribution of two skyrmions with opposite winding directions (Figure 2(f)) reveals the cooperative mechanism between magnetocrystalline anisotropy and external magnetic fields.

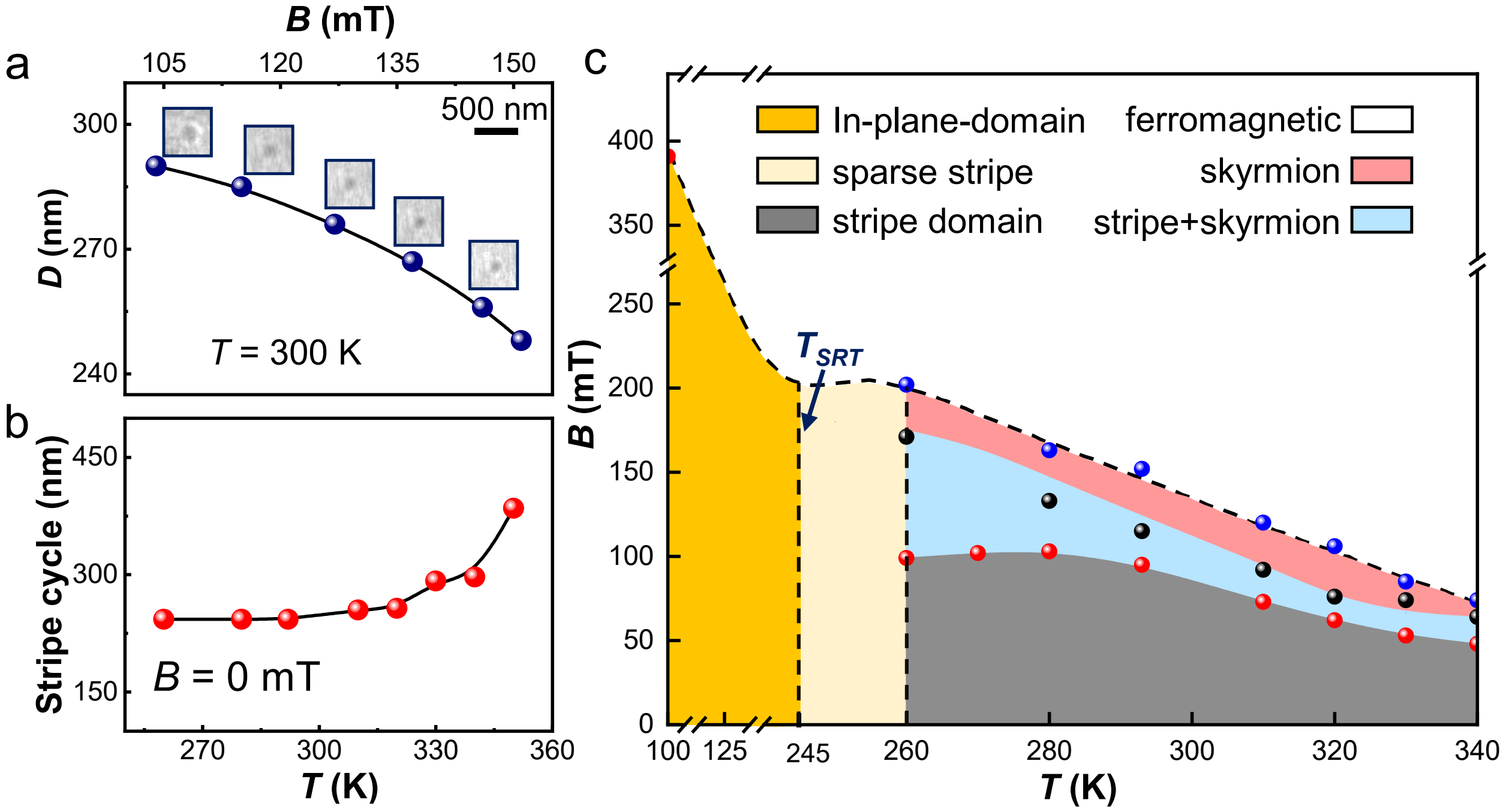

**Figure 3.** Evolution of topological magnetic structures in $Mn_{1.9}Co_{0.1}Sb$ layered system under external field and temperature regulation. a) Skyrmion diameter as a function of the magnetic field at $T$ = 300 K. b) Stripe cycle as a function of the temperature at zero field. c) Magnetic phase diagram of magnetic domains as a function of temperature and magnetic field.

To systematically investigate how the magnetic structure changes in size in response to an applied magnetic field to achieve a new energy balance, the skyrmion diameter as a function of the applied magnetic field strength at 300 K is shown in **Figure 3**(a). Initially, magnetic skyrmions are generated by increasing the magnetic field from zero. As the magnetic field increases, the diameter of the skyrmions gradually decreases, ranging from approximately 250 nm to 290 nm. When the magnetic field reaches a certain threshold, the external field gradually dominates, altering the internal energy balance of the particles. To minimize the total system energy, the skyrmions adjust their size and shape in response to changes in the external field, leading to an accelerated decrease in skyrmion size and ultimately annihilating at 150 mT. Additionally, temperature also affects the size of the initial ground state stripe domains at zero field. Influenced by thermal excitation and magnetic crystalline anisotropy, during a zero-field heating process from 260 K to 350 K, thermal fluctuations increase, and the effect of perpendicular magnetic anisotropy within the material weakens. Consequently, the periodic width of the ground state stripe domains gradually increases, from approximately 240 nm to 390 nm, as shown in Figures 3(b) and **Figure 4**.

Based on our observations, we have compiled a comprehensive phase diagram illustrating the evolution of magnetic domain states as a function of temperature and magnetic field, as shown in Figure 3(c). The colored dots represent the threshold magnetic fields for the transitions from stripe domains to ferromagnetic states or skyrmions, and from skyrmions to ferromagnetic states during the field increase process. The colored regions represent the states of magnetic structures such as stripe domains, multi-stripes states, and skyrmions during the field increase process. Within the temperature range of 100 K to 250 K, the magnetic domains at zero field are all in-plane stripe domains. As the magnetic field gradually increases, these domains transition into a ferromagnetic state. Between 250 K and 260 K, a multi-domain state coexisting with out-of-plane and in-plane stripe domains is observed at zero field. In the temperature range of 260 K to 340 K, the initial state is stripe domains. As the magnetic field increases, some of these stripe domains shrink and transform into skyrmions. Eventually, all domains transition into skyrmions. With further increase in the magnetic field, the system reaches saturation and transforms into a ferromagnetic state.

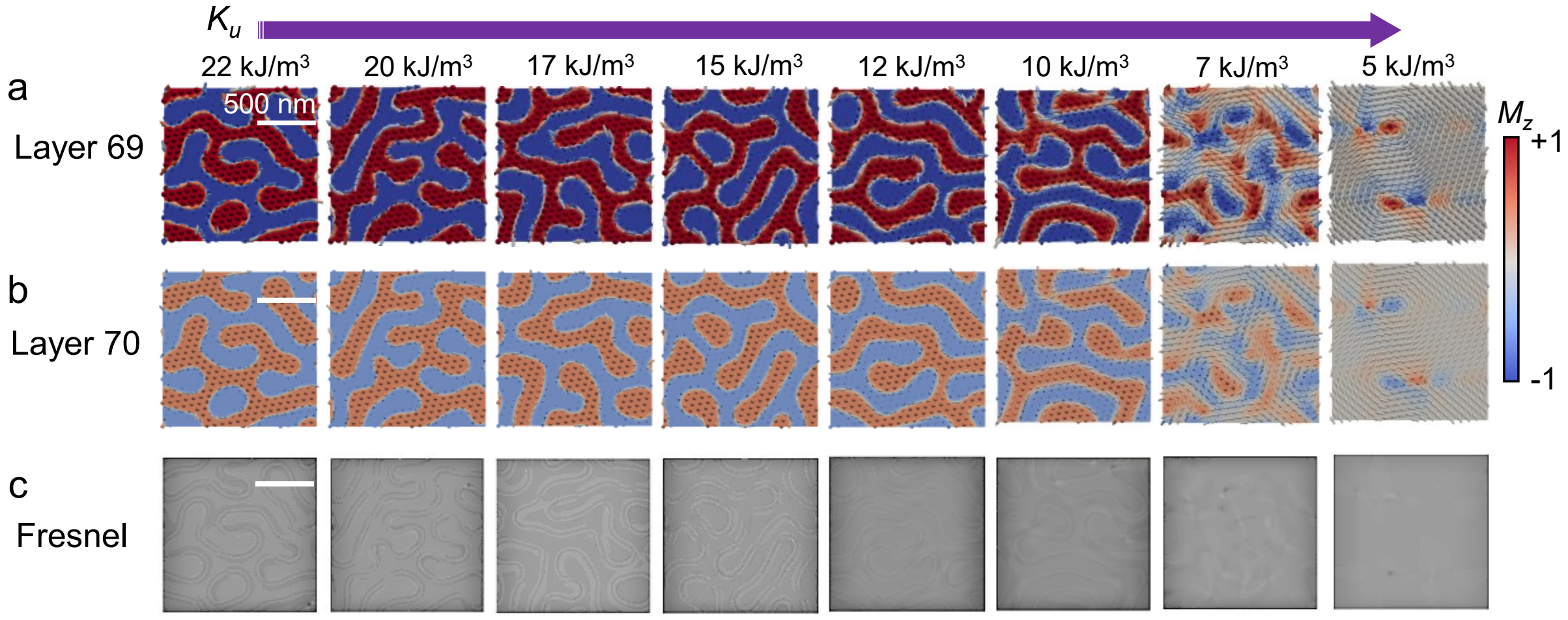

**Figure 4**. Simulated magnetization distributions and simulated Fresnel defocused images of • $Mn_{1.9}Co_{0.1}Sb$. a) Micromagnetic simulations of intermediate-layer 69, b) layer 70 and c) Fresnel simulation evolution of zero-field stripe domains by decreasing the magnetocrystalline anisotropy $K_u$.

For this system, the calculated antiparallel magnetic moment of 0.9 μB /Mn agrees with our experimental result 0.97 μB/(Mn/Co) and the previous neutron diffraction result, [42] which confirm the antiferromagnetic aliments of interlayer Mn atoms. Therefore, in our simulations, adjacent Mn layers were set to antiparallel alignment. Figure 4 presents simulated magnetization distributions and simulated Fresnel defocused images based on the magnetic parameters of $Mn_{1.9}Co_{0.1}Sb$. The results clearly demonstrate that the magnetization in layer 69 and layer 70 exhibits distinct magnitudes and antiparallel alignment, consistent with characteristic ferrimagnetic ordering. Furthermore, at a magnetic anisotropy constant of $K_u$ = 22 kJ/m³, the simulation results (Figure S7) show that the remanence of its M-H magnetization curve remains close to zero. This is primarily attributed to the spontaneous formation of stripe domains in the material under zero field. This multi-domain structure causes the internal magnetization directions to cancel each other out macroscopically, resulting in very low remanence. As $K_u$ is progressively reduced, the width of these stripe domains correspondingly decreases, ultimately evolving into in-plane magnetized domains.

### 2.3. Magnetic Excitations of Dipolar Skyrmions in Ferrimagnetic Lamella

We constructed a $Mn_{1.9}Co_{0.1}Sb$ ferrimagnetic model based on antiferromagnetic coupling consistent with synthetic ferrimagnetic/antiferromagnetic systems. This system featuring in-plane ferromagnetic coupling, interlayer antiferromagnetic arrangement, and non-equivalent magnetic moments between adjacent layers, to investigate the skyrmion dynamics. **Figure 5**(a) displays the schematic skyrmion configuration in this ferrimagnetic system under the influence of demagnetization energy, revealing antiparallel-aligned Bloch-type skyrmions in intermediate layers that determine the net magnetization, while Néel-type skyrmions dominate the top and bottom layers. [54] Figures 5(b)-(g) display the in-plane magnetization distribution between two adjacent layers at the top, intermediate, and bottom levels, all of which exhibit an antiparallel alignment, as implemented in our simulation mode.  Magnified views of the fine structure demonstrate the crucial role of dipolar interactions in stabilizing such configurations. The ferrimagnetic model possesses dipolar character and intrinsic interatomic exchange coupling, which is contrast to synthetic ferrimagnetic/antiferromagnetic systems, that exhibit interlayer antiferromagnetic coupling via the RKKY interaction and purely DMI-stabilized Néel-type skyrmions. [55-58] In synthetic ferrimagnets /antiferromagnets, the antiferromagnetic coupling could contribute only several tens of GHz dynamics because of weak RKKY interactions. [55-58] By contrast, the unique topological structure and strong interatomic layer antiparallel exchange interactions suggest that ferrimagnetic skyrmions may exhibit sub-THz high-frequency dynamics. To investigate the influence of the exchange interaction on the skyrmion dynamics, we examined the magnetic absorption spectrum of skyrmions by applying a sinc-function pulse field $B_x = B_0\mathrm{sinc}(2\pi ft)$ (where $h_0$ = 1 mT and $2\pi f$ =1000 GHz) combined with an external field $B_z$ = 20 mT to the stable skyrmion state. [22, 30] The magnetic absorption spectrum of skyrmions was then calculated by recording the spatially averaged magnetization evolution and performing a Fourier transformation to obtain the dynamic susceptibility $\chi_{xx}$.

Through comparative analysis of the frequency-$\chi_{xx}$ curves between the saturated state and skyrmion state (Figure 5(h)), we identified the characteristic high-frequency resonance position of skyrmions, the orange labels 1-8 indicate distinct resonance frequencies of the skyrmion state, with the brown dashed line specifically marking the high-frequency resonance peak at 186.8 GHz (marked as peak 7 in the brown curve). To elucidate the influence of exchange coupling $\sigma$ on skyrmion excitation modes and their dynamic properties, we systematically varied $\sigma$ from $0.75\times10^{-12}$ J/m to $2.50\times10^{-12}$ J/m under fixed external field conditions (see Figure S8). Rotation modes can be excited when the in-plane ac field is applied, and the breathing mode is preferred when the ac field is perpendicular to the system. Inset presents the spatial power spectra for $\sigma$ = $1.0\times10^{-12}$ J/m. The spatial power spectrum is typically employed to characterize the distribution of magnetization fluctuations in space, which agrees with the simulation results showing rotational mode at eigenfrequency peak position 7 in Supplementary Video S1.

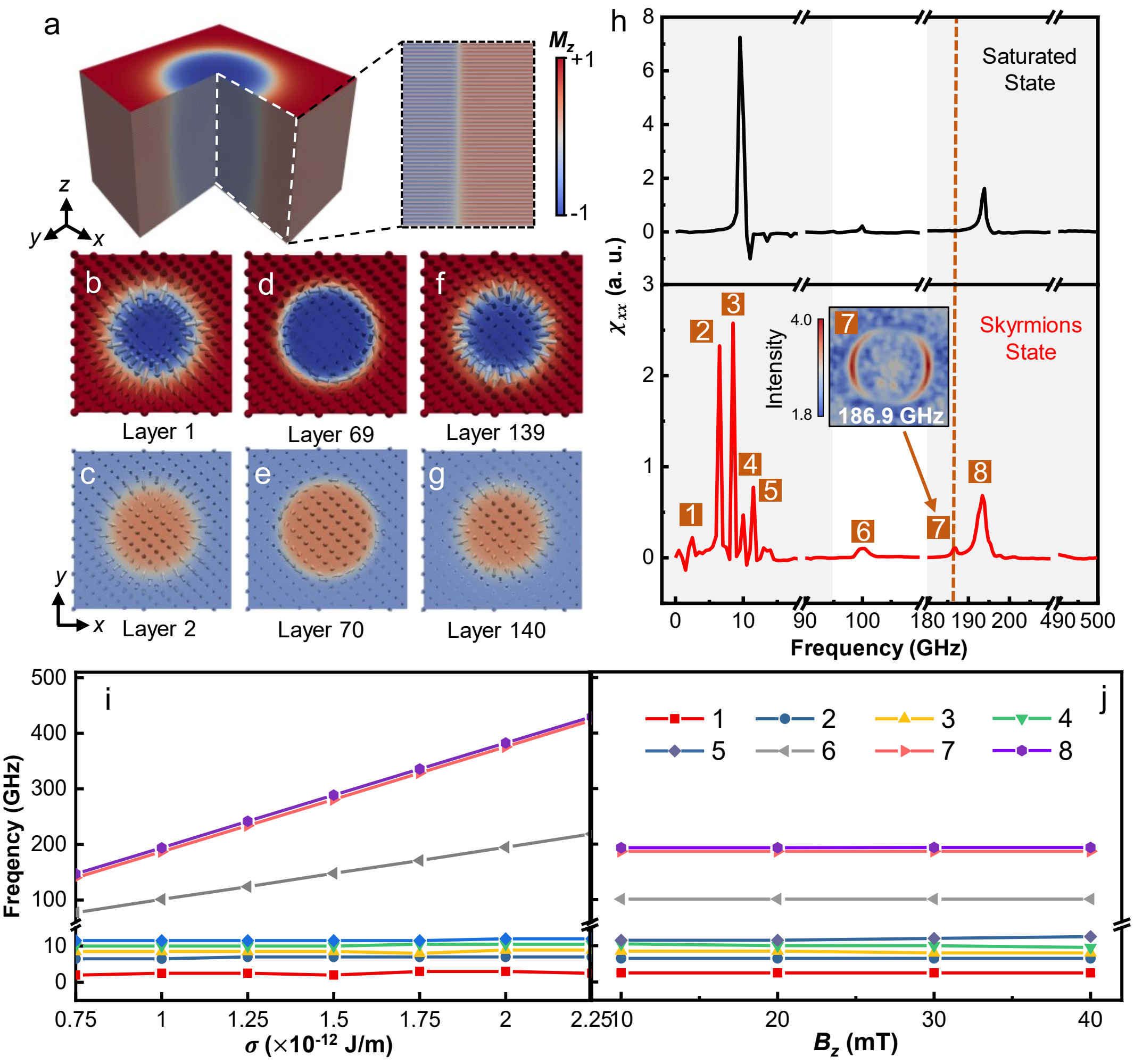

**Figure 5**. Simulated magnetic excitations of $Mn_{1.9}Co_{0.1}Sb$. a) Ferrimagnetic structure diagram and side view of $Mn_{1.9}Co_{0.1}Sb$; b)-g) represent the in-plane magnetization distributions for layer 1, layer 2, layer 69, layer 70, layer 139, and layer 140, respectively. h) The comparative frequency-dynamical susceptibility between the saturated state and skyrmion state at $\sigma = 1.0 \times 10^{-12}$ J/m. The gray and white background regions demarcate the low/medium/high frequency ranges, respectively. i) The resonance frequencies of different excitation modes as a function of the $\sigma$. j) The resonance frequencies of different excitation modes as a function of the $B_z$.

To classify these resonance modes, we plot the resonance frequency as a function of coupling strength $\sigma$ in Figure 5(i) and Figure S8. The results show that the peak positions in different frequency bands exhibit significant differences in their responses to $\sigma$, revealing a hierarchical regulation mechanism of the coupling strength on the multi-scale dynamics of the system. In the low-frequency band (peak 1-5), the frequency remains stable. The frequency stability in the low-frequency regime (peaks 1-5) stems from the $\sigma$ insensitivity of global cooperative motion, which renders the system frequency unaffected by variations in coupling strength. Concurrently, characteristic peaks in the low-wavevector region of the spatial power spectrum exhibit significant sharpening. The mid-frequency band (peak 6) demonstrates linear frequency growth with increasing $\sigma$, while the high-frequency band (peaks 7-8) displays a steeper slope of frequency increase. This dynamical mechanism originates from higher-order collective excitations through suppression of local fluctuations. Finally, we also discussed the distribution of $\chi_{xx}$ for different external fields in the low-frequency region and high-frequency region when the fixed $\sigma = 1.0 \times 10^{-12}$ J/m (as shown in Figure 5(j) and Figure S9). It shows that in the low-frequency region, as the external field increases, the resonance frequency decreases slightly, while the high-frequency region shows a stable state. This is because the resonance in the high-frequency region is dominated by the exchange interaction. Since $\sigma$ is fixed, the contribution of the exchange field is much greater than the influence of the external field, so changes in the external field have little effect on the high-frequency region. As a ferrimagnetic material, $Mn_{1.9}Co_{0.1}Sb$ demonstrates remarkable advantages in magnetic memory applications, including high resonant frequencies and ultrafast response, owing to its ferrimagnetic ordering and strong exchange coupling.

## 3. Conclusions

In summary, this study successfully synthesized $Mn_{1.9}Co_{0.1}Sb$ single crystals via the melt-cooling method, which were theoretically predicted to host magnetic skyrmions. Under perpendicular magnetic fields, we observed stable magnetic skyrmions at room temperature using LTEM and revealed the field-driven transformation process from stripe domains to magnetic skyrmions. Based on the magnetic field-temperature phase diagram, we systematically established the evolution principles of topological spin structures in $Mn_{1.9}Co_{0.1}Sb$. Furthermore, investigations of terahertz-frequency magnetic excitations in ferrimagnetic skyrmions confirm their potential as promising candidates for ultrafast magnetic memory devices. This work not only delivers theoretical foundations, key experimental evidence, and micromagnetic simulation validations but also establishes $Mn_{1.9}Co_{0.1}Sb$ as a pivotal material platform for topological magnetism research, creating new pathways for developing advanced spintronic devices.

## 4. Experimental Section

*Sample Preparation*: $Mn_{1.9}Co_{0.1}Sb$ single crystals were grown by the melt cooling method with stoichiometric manganese sheets (99.9%), cobalt particles (99.99%), and antimony chunks (99.99%) mixed as a molar ratio of 1.9:0.1:1. The mixed raw materials were loaded in an alumina crucible inside evacuated fused quartz tubes. The tube was heated to 1010 ℃, held for 1 day to make the mixture react sufficiently, and was slowly cooled down to 860 ℃ by a ratio of 1 ℃ per hour, followed by subsequent quenching to obtain $Mn_{1.9}Co_{0.1}Sb$ single crystals while avoiding impurities.

*Fabrication of $Mn_{1.9}Co_{0.1}Sb$ Lamellas:* The 140 nm-thick $Mn_{1.9}Co_{0.1}Sb$ nanostructured lamellas were fabricated from a bulk single crystal using a standard lift-out method, with a focused ion beam and scanning electron microscopy dual-beam system (Helios Nanolab 600i, FEI)[50].

*TEM Measurements*: In situ Fresnel imaging was used in Lorentz TEM (Talos F200X, FEI) with an acceleration voltage of 200 kV to investigate magnetic domains in the $Mn_{1.9}Co_{0.1}Sb$ lamellas. The TEM holder (model 636.6, Gatan) could support the varying temperature measurements.

*Micromagnetic Simulations:* The micromagnetic simulations for magnetic domains in 140 nm-thick lamellas were performed using MuMax3, [59] where the antiferromagnetic coupling between two magnetic sublattices was modeled through alternating layered structures with negative exchange interactions. It was considered in the Hamiltonian exchange interaction ($A$) energy, uniaxial magnetic anisotropy ($K_u$) energy, Zeeman energy, and dipole-dipole interaction energy. The modeling approach involved programmatically defining odd-numbered layers as Mn(I) sublattices and even-numbered layers as Mn(II) sublattices, creating an alternating Mn(I)-Mn(II)-Mn(I)... stacking sequence. The simulation parameters were configured based on the distinct magnetic moments of Mn(I) and Mn(II) in $Mn_{1.9}Co_{0.1}Sb$, with anisotropy constants $K_u = 22$ kJ·m$^{-3}$, saturation magnetizations $M_s$ (I) = 31.1 kA·m$^{-1}$ and $M_s$ (II) = 65.2 kA·m$^{-1}$, the exchange constant was set to $A = 4$ pJ·m$^{-1}$. The computational cell size was 4×4×1 nm$^3$.

**Supporting Information**

Supporting Information is available from the Wiley Online Library or from the author.

**Acknowledgments**

This work was supported by the National Key R&D Program of China, Grant No.

2022YFA1403603; the Scientific Research Innovation Capability Support Project for Young Faculty, Grant No. SRICSPYF-BS2025073; the Natural Science Foundation of China, Grants No. 52325105, 12422403, U24A6001, 12174396, 12104123, and 12241406; the Anhui Provincial Natural Science Foundation, Grant No. 2308085Y32 and 2508085MA018; the Natural Science Project of Colleges and Universities in Anhui Province, Grants No. 2022AH030011 and 2024AH030046; the Fundamental Research Funds for the Central Universities (Grant No. 104972025KFYjc0089); the Strategic Priority Research Program of Chinese Academy of Sciences, Grant No. XDB33030100; CAS Project for Young Scientists in Basic Research, Grant No. YSBR-084; Systematic Fundamental Research Program Leveraging Major Scientific and Technological Infrastructure, Chinese Academy of Sciences, Grant No. JZHKYPT-2021-08; Anhui Province Excellent Young Teacher Training Project, Grant No. YQZD2023067; the 2024 Project of GDRCYY (No. 217, Yaodong Wu); and the China Postdoctoral Science Foundation, Grant No. 2023M743543. We thank the staff members of the Physical Property Measurement System (https://cstr.cn/31125.02.SHMFFPPMS) at the Steady High Magnetic Field Facility, CAS (https://cstr.cn/31125.02.SHMFF), for providing technical support and assistance in data collection and analysis.

**Author Contributions**

H. Z. and Y. K. contributed equally to this work. J. T., Y. K., Y.W., and H. D. supervised the project. J. T., Y. K., and H. Z. conceived the idea and designed the experiments. Y. K. and L. C. synthesized the $Mn_{1.9}Co_{0.1}Sb$ single crystals. Y. K. and H. Z. measured their crystals and macroscopic magnetism. H. Z. fabricated the $Mn_{1.9}Co_{0.1}Sb$ nanostructured cells. H. Z., J. T., Y.W., Y. P., S. Q., J. J., and Y. Z. performed the TEM measurements. H. Z. J. T., W. W., and L. K. performed the simulations. H. Z., Y. K., Y. W., H. D., and J. T. wrote the manuscript with input from all authors. All authors discussed the results and contributed to the manuscript.

## Conflict of Interest

The authors declare no competing financial interest.

## References

[1] N. Nagaosa, Y. Tokura, "Topological Properties and Dynamics of Magnetic Skyrmions", *Nat. Nanotechnol.* **2013**, *8*, 899.

[2] A. Fert, V. Cros, J. Sampaio, "Skyrmions on the Track", *Nat. Nanotechnol.* **2013**, *8*, 152.

[3] Y. Zhou, "Magnetic Skyrmions: Intriguing Physics and New Spintronic Device Concepts", *Natl. Sci. Rev.* **2019**, *6*, 210.

[4] A. Fert, N. Reyren, V. Cros, "Magnetic Skyrmions: Advances in Physics and Potential Applications", *Nat. Rev. Mater.* **2017**, *2*, 17031.

[5] S. Li, X. Wang, T. Rasing, "Magnetic Skyrmions: Basic Properties and Potential Applications", *Interdiscip. Mater.* **2023**, *2*, 260.

[6] W. Li, C. Jin, R. Che, W. Wei, L. Lin, L. Zhang, H. Du, M. Tian, J. Zang, "Emergence of Skyrmions from Rich Parent Phases in the Molybdenum Nitrides", *Phys. Rev. B* **2016**, *93*, 060409.

[7] B. Göbel, I. Mertig, O. A. Tretiakov, "Beyond Skyrmions: Review and Perspectives of Alternative Magnetic Quasiparticles", *Phys. Rep.* **2021**, *895*, 1.

[8] T. Schulz, R. Ritz, A. Bauer, M. Halder, M. Wagner, C. Franz, C. Pfleiderer, K. Everschor, M. Garst, A. Rosch, "Emergent Electrodynamics of Skyrmions in a Chiral Magnet", *Nat. Phys.* **2012**, *8*, 301.

[9] S. Mühlbauer, F. Jonietz, C. Pfleiderer, A. Rosch, A. Neubauer, R. Georgii, P. Böni, "Skyrmion Lattice in a Chiral Magnet", *Science* **2009**, *323*, 915.

[10] X. Z. Yu, Y. Onose, N. Kanazawa, J. H. Park, J. H. Han, Y. Matsui, N. Nagaosa, Y. Tokura, "Real-Space Observation of a Two-Dimensional Skyrmion Crystal", *Nature* **2010**, *465*, 901.

[11] X. Z. Yu, N. Kanazawa, Y. Onose, K. Kimoto, W. Z. Zhang, S. Ishiwata, Y. Matsui, Y. Tokura, "Near Room-Temperature Formation of a Skyrmion Crystal in Thin-Films of the Helimagnet FeGe", *Nat. Mater.* **2010**, *10*, 106.

[12] J. Tang, Y. Wu, W. Wang, L. Kong, B. Lv, W. Wei, J. Zang, M. Tian, H. Du, "Magnetic Skyrmion Bundles and Their Current-Driven Dynamics", *Nat. Nanotechnol.* **2021**, *16*, 1086.

[13] X. Z. Yu, K. Nakajima, K. Shibata, N. Kanazawa, T. Arima1, N. Nagaosa1, Y. Tokura,, "Motion Tracking of 80-nm-Size Skyrmions Upon Directional Current Injections", *Sci. Adv.* **2020**, *6*, eaaz9744.

[14] Y. Tokunaga, X. Z. Yu, J. S. White, H. M. Rønnow, D. Morikawa, Y. Taguchi, Y. Tokura, "A New Class of Chiral Materials Hosting Magnetic Skyrmions Beyond Room Temperature", *Nat. Commun.* **2015**, *6*, 7638.

[15] W. Wang, D. Song, W. Wei, P. Nan, S. Zhang, B. Ge, M. Tian, J. Zang, H. Du, "Electrical Manipulation of Skyrmions in a Chiral Magnet", *Nat. Commun.* **2022**, *13*, 1593.

[16] W. Kang, Y. Huang, X. Zhang, Y. Zhou, W. Zhao, "Skyrmion-Electronics: An Overview and Outlook", *Proc. IEEE* **2016**, *104*, 2040.

[17] S. Woo, K. Litzius, B. Krüger, M.-Y. Im, L. Caretta, K. Richter, M. Mann, A. Krone,

R. M. Reeve, M. Weigand, P. Agrawal, I. Lemesh, M.-A. Mawass, P. Fischer, M. Kläui, G. S. D. Beach, "Observation of Room-Temperature Magnetic Skyrmions and Their Current-Driven Dynamics in Ultrathin Metallic Ferromagnets", *Nat. Mater.* **2016**, *15*, 501.

[18] Y. Wu, J. Jiang, W. Wang, L. Kong, S. Wang, M. Tian, H. Du, J. Tang, "Skyrmion Sliding Switch in a 90 nm-Wide Nanostructured Chiral Magnet", *Nano Lett.* **2025**, *25*, 7012.

[19] A. M. Cook, "Quantum Skyrmion Hall Effect", *Phys. Rev. B* **2024**, *109*, 155123.

[20] W. Jiang, X. Zhang, G. Yu, W. Zhang, X. Wang, M. Benjamin Jungfleisch, John E. Pearson, X. Cheng, O. Heinonen, K. L. Wang, Y. Zhou, A. Hoffmann, Suzanne G. E. te Velthuis, "Direct Observation of the Skyrmion Hall Effect", *Nat. Phys.* **2016**, *13*, 162.

[21] D. Maccariello, W. Legrand, N. Reyren, K. Garcia, K. Bouzehouane, S. Collin, V. Cros, A. Fert, "Electrical Detection of Single Magnetic Skyrmions in Metallic Multilayers at Room Temperature", *Nat. Nanotechnol.* **2018**, *13*, 233.

[22] W. Wang, M. Beg, B. Zhang, W. Kuch, H. Fangohr, "Driving Magnetic Skyrmions with Microwave Fields", *Phys. Rev. B* **2015**, *92*, 020403.

[23] Y. Okamura, F. Kagawa, M. Mochizuki, M. Kubota, S. Seki, S. Ishiwata, M. Kawasaki, Y. Onose, Y. Tokura, "Microwave Magnetoelectric Effect Via Skyrmion Resonance Modes in a Helimagnetic Multiferroic", *Nat. Commun.* **2013**, *4*, 2391.

[24] X. Liang, X. Zhang, L. Shen, J. Xia, M. Ezawa, X. Liu, Y. Zhou, "Dynamics of Ferrimagnetic Skyrmionium Driven by Spin-Orbit Torque", *Phys. Rev. B* **2021**, *104*, 174421.

[25] H. Wu, F. Groß, B. Dai, D. Lujan, S. A. Razavi, P. Zhang, Y. Liu, K. Sobotkiewich, J.

Förster, M. Weigand, G. Schütz, X. Li, J. Gräfe, K. L. Wang, "Ferrimagnetic Skyrmions in Topological Insulator/Ferrimagnet Heterostructures", *Adv. Mater.* **2020**, *32*, 2003380.

[26] T. Xu, Y. Zhang, Z. Wang, H. Bai, C. Song, J. Liu, Y. Zhou, S.-G. Je, A. T. N'Diaye, M.-Y. Im, R. Yu, Z. Chen, W. Jiang, "Systematic Control of Ferrimagnetic Skyrmions Via Composition Modulation in Pt/$Fe_{1-x}Tb_x$/Ta Multilayers", *ACS Nano* **2023**, *17*, 7920.

[27] L. Bo, X. Zhang, M. Mochizuki, X. Zhang, "Suppression of the Skyrmion Hall Effect in Synthetic Ferrimagnets with Gradient Magnetization", *Phys. Rev. Res.* **2024**, *6*, 023199.

[28] R. C. Silva, R. L. Silva, J. C. Moreira, W. A. Moura-Melo, A. R. Pereira, "Channeling Skyrmions: Suppressing the Skyrmion Hall Effect in Ferrimagnetic Nanostripes", *J. Appl. Phys.* **2024**, *135*, 183902.

[29] J. Luo, J. Xia, X. Liang, Y. Zhou, L. Li, L. Gu, G. Zhao, "Dynamics of Skyrmions in Synthetic Ferrimagnetic Materials", *Phys. Rev. B* **2024**, *110*, 214409.

[30] L. Xing, D. Hua, W. Wang, "Magnetic Excitations of Skyrmions in Antiferromagnetic-Exchange Coupled Disks", *J. Appl. Phys.* **2018**, *124*, 123904.

[31] Y. Liu, T. T. Liu, Z. Jin, Z. P. Hou, D. Y. Chen, Z. Fan, M. Zeng, X. B. Lu, X. S. Gao, M. H. Qin, J. M. Liu, "Spin-Wave-Driven Skyrmion Dynamics in Ferrimagnets: Effect of Net Angular Momentum", *Phys. Rev. B* **2022**, *106*.

[32] Y. Liu, T. T. Liu, Z. P. Hou, D. Y. Chen, Z. Fan, M. Zeng, X. B. Lu, X. S. Gao, M. H. Qin, J. M. Liu, "Dynamics of Hybrid Magnetic Skyrmion Driven by Spin–Orbit Torque in Ferrimagnets", *Appl. Phys. Lett.* **2023**, *122*.

[33] S. Woo, K. M. Song, X. Zhang, Y. Zhou, M. Ezawa, X. Liu, S. Finizio, J. Raabe, N. J. Lee, S.-I. Kim, S.-Y. Park, Y. Kim, J.-Y. Kim, D. Lee, O. Lee, J. W. Choi, B.-C. Min,

H. C. Koo, J. Chang, "Current-Driven Dynamics and Inhibition of the Skyrmion Hall Effect of Ferrimagnetic Skyrmions in Gdfeco Films", *Nat. Commun.* **2018**, *9*, 959.

[34] S. Mallick, Y. Sassi, N. F. Prestes, S. Krishnia, F. Gallego, L. M. Vicente Arche, T. Denneulin, S. Collin, K. Bouzehouane, A. Thiaville, R. E. Dunin-Borkowski, V. Jeudy, A. Fert, N. Reyren, V. Cros, "Driving Skyrmions in Flow Regime in Synthetic Ferrimagnets", *Nat. Commun.* **2024**, *15*, 8472.

[35] L. Caretta, M. Mann, F. Büttner, K. Ueda, B. Pfau, C. M. Günther, P. Hessing, A. Churikova, C. Klose, M. Schneider, D. Engel, C. Marcus, D. Bono, K. Bagschik, S. Eisebitt, G. S. D. Beach, "Fast Current-Driven Domain Walls and Small Skyrmions in a Compensated Ferrimagnet", *Nat. Nanotechnol.* **2018**, *13*, 1154.

[36] M. I. Bartashevich, T. Goto, N. V. Baranov, V. S. Gaviko, "Volume Magnetostriction at the AF–FRI Metamagnetic Transition in the Itinerant-Electron System $Mn_{2-x}T_xSb$ (T=Co, Cr)", *Physica B* **2004**, *351*, 71.

[37] S. C. Ma, D. Hou, Y. Y. Gong, L. Y. Wang, Y. L. Huang, Z. C. Zhong, D. H. Wang, Y. W. Du, "Giant Magnetocaloric and Magnetoresistance Effects in Ferrimagnetic $Mn_{1.9}Co_{0.1}Sb$ Alloy", *Appl. Phys. Lett.* **2014**, *104*, 022410.

[38] T. I. T. Kanomata, Y. Hasebe, H. Yoshida, T. Kaneko, "Pressure Effect on Magnetic Transition Temperature of $Mn_{2-x}Co_xSb$", *J. Magn. Magn. Mater.* **1990**, *90*, 719.

[39] T. Kanomata, H. Ido, "Magnetic Transitions in $Mn_{2-x}M_xSb$ (M=3d Metals)", *J. Appl. Phys.* **1984**, *55*, 2039.

[40] S. Ma, G. Yu, C. Qiu, J. Liu, Z. Zhang, X. Luo, C. Chen, C. Fang, Y. Yuan, Z. Zhong, "Topological Hall Effect Associated with Spin Reorientation Transition in Tetragonal $Mn_{1.9}Co_{0.1}Sb$ Ferrimagnet", *J. Alloys Compd.* **2022**, *921*.

[41] L. Heaton, "The Crystal Structure of $Mn_2Sb$", *Acta Crystallogr.* **1955**, *8*, 207.

[42] M. K. Wilkinson, C. G. Shull, "The Magnetic Structure of $Mn_2Sb$", *J. Phys. Chem. Solids* **1957**, *2*, 289.

[43] J. S. Wilden, A. Hoser, M. Chikovani, J. Perßon, J. Voigt, K. Friese, A. Grzechnik, "Magnetic Transitions in the Co-Modified $Mn_2Sb$ System", *Inorganics* **2018**, *6*, 113.

[44] Y. Q. Zhang, Z. D. Zhang, "Giant Magnetoresistance and Magnetocaloric Effects of the $Mn_{1.82}V_{0.18}Sb$ Compound", *J. Alloys Compd.* **2004**, *365*, 35.

[45] L. Caron, X. F. Miao, J. C. P. Klaasse, S. Gama, E. Brück, "Tuning the Giant Inverse Magnetocaloric Effect in $Mn_{2-x}Cr_xSb$ Compounds", *Appl. Phys. Lett.* **2013**, *103*, 112404.

[46] Y. Cao, K. Xu, Z. Li, Y. Zhang, X. He, Y. Kang, W. Sun, T. Gao, Z. Qian, C. Liu, M. Ye, C. Jing, "Interplay between Spin Reorientation and Magnetoelastic Transitions, and Anisotropic Magnetostriction in the $Mn_{1.95}Cr_{0.05}Sb$ Single Crystal", *J. Magn. Magn. Mater.* **2019**, *487*, 165315.

[47] Y. Matsumoto, K. Matsubayashi, Y. Uwatoko, M. Hiroi, Y. Mitsui, K. Koyama, "Magnetic Properties of $Mn_{1.9}Cu_{0.1}Sb$ under High Pressure", *AIP Conf. Proc.* **2016**, *1763*, 020005.

[48] Y. Li, M. R. U. Nabi, H. Park, Y. Liu, S. Rosenkranz, A. K. Petford‐Long, J. Hu, S. G. E. t. Velthuis, C. Phatak, "Observation of Topological Spin Textures in Ferrimagnetic $Mn_{2-x}Zn_xSb$", *Small* **2025**, *21*, 2406299.

[49] J. Tang, Y. Wu, L. Kong, W. Wang, Y. Chen, Y. Wang, Y. Soh, Y. Xiong, M. Tian, H. Du, "Two-Dimensional Characterization of Three-Dimensional Magnetic Bubbles in $Fe_3Sn_2$ Nanostructures", *Natl. Sci. Rev.* **2021**, *8*, nwaa200.

[50] L. Kong, J. Tang, Y. Wu, W. Wang, J. Jiang, Y. Wang, J. Li, Y. Xiong, S. Wang, M. Tian, H. Du, "Diverse Helicities of Dipolar Skyrmions", *Phys. Rev. B* **2024**, *109*, 014401.

[51] M. Hassan, S. Koraltan, A. Ullrich, F. Bruckner, R. O. Serha, K. V. Levchenko, G. Varvaro, N. S. Kiselev, M. Heigl, C. Abert, D. Suess, M. Albrecht, "Dipolar Skyrmions and Antiskyrmions of Arbitrary Topological Charge at Room Temperature", *Nat. Phys.* **2024**, *20*, 615.

[52] J. Tang, J. Jiang, Y. Wu, L. Kong, Y. Wang, J. Li, Y. Soh, Y. Xiong, S. Wang, M. Tian, H. Du, "Creating and Deleting a Single Dipolar Skyrmion by Surface Spin Twists", *Nano Lett.* **2024**, *25*, 121.

[53] K. Ishizuka, B. Allman, "Phase Measurement of Atomic Resolution Image Using Transport of Intensity Equation", *Microscopy* **2005**, *54*, 191.

[54] J. Tang, L. Kong, Y. Wu, W. Wang, Y. Chen, Y. Wang, J. Li, Y. Soh, Y. Xiong, M. Tian, H. Du, "Target Bubbles in $Fe_3Sn_2$ Nanodisks at Zero Magnetic Field", *ACS Nano* **2020**, *14*, 10986.

[55] X. Chen, C. Zheng, S. Zhou, Y. Liu, Z. Zhang, "Ferromagnetic Resonance Modes of a Synthetic Antiferromagnet at Low Magnetic Fields", *J. Phys.: Condens. Matter* **2022**, *34*, 015802.

[56] H. J. Waring, N. A. B. Johansson, I. J. Vera-Marun, T. Thomson, "Zero-Field Optic Mode Beyond 20 GHz in a Synthetic Antiferromagnet", *Phys. Rev. Appl.* **2020**, *13*, 034035.

[57] M. Lonsky, A. Hoffmann, "Coupled Skyrmion Breathing Modes in Synthetic Ferri- and Antiferromagnets", *Phys. Rev. B* **2020**, *102*, 104403.

[58] L. Qiu, L. Shen, X. Zhang, Y. Zhou, G. Zhao, W. Xia, H.-B. Luo, J. P. Liu, "Interlayer

Coupling Effect on Skyrmion Dynamics in Synthetic Antiferromagnets", *Appl. Phys. Lett.* **2021**, *118*, 082403.

[59] A. Vansteenkiste, J. Leliaert, M. Dvornik, M. Helsen, F. Garcia-Sanchez, B. Van Waeyenberge, "The Design and Verification of Mumax3", *AIP Adv.* **2014**, *4*, 107133.